\documentclass[conference]{IEEEtran}
\IEEEoverridecommandlockouts
\usepackage{cite}
\usepackage{hyperref}
\usepackage{amsmath,amssymb,amsfonts}
\usepackage{algpseudocode}
\usepackage{algorithm}
\usepackage{graphicx}
\usepackage{textcomp}
\usepackage{tabularx}
\usepackage{xcolor}
\usepackage{url}
\usepackage{tikz}
\usepackage{csquotes}
\usepackage{listings, lstautogobble}

\newcommand\copyrighttext{%
  \footnotesize This is the author’s version of the work. It is posted here for your personal use. Not for redistribution.
The definitive Version of Record is available in the proceedings of IEEE 45th International Conference on Distributed Computing Systems Workshops (ICDCSW), 2025 (DOI: 10.1109/ICDCSW63273.2025.00107).}
\newcommand\copyrightnotice{%
\begin{tikzpicture}[remember picture,overlay]
\node[anchor=south,yshift=10pt] at (current page.south) {\fbox{\parbox{\dimexpr\textwidth-\fboxsep-\fboxrule\relax}{\copyrighttext}}};
\end{tikzpicture}%
}

\begin{document}

\title{A Self-Adaptive Extensible CEP framework for the  Cloud-Edge Continuum

}

\author{
	\IEEEauthorblockN{ Olaf Markus Link}
	\IEEEauthorblockA{\textit{University of Potsdam} \\
		Potsdam, Germany \\
		olaf.markus.link@uni-potsdam.de}
	\and
\IEEEauthorblockN{Sukanya Bhowmik}
	\IEEEauthorblockA{\textit{University of Potsdam} \\
		Potsdam, Germany \\
		sukanya.bhowmik@uni-potsdam.de}
        }



\maketitle
\thispagestyle{plain}
\pagestyle{plain}

\begin{abstract}

Traditional complex event processing  (CEP) systems focus on extracting information out of simple  events in the input event streams to detect  complex event patterns.  In the context of CEP, a computing unit that executes a complex data transformation task is called an operator. Typically, a CEP system consists of an operator graph that could be distributed over the cloud-edge continuum. However, an operator graph distributed over  heterogeneous nodes comes with its own set of challenges. In fact, the nodes of an operator graph are typically subjected to dynamically changing workloads. Unbalanced load distributions across  heterogeneous nodes may lead to   processing overloads in individual operators that do not have enough resources to properly  adapt to an increasing workload. As a result, a framework that can detect and adapt to dynamically changing workload in real time across the various nodes of the cloud-edge continuum becomes paramount. To this end, the goal of this work is to propose a new extensible self-adaptive CEP framework with load  regulatory features that help operators continue working even in face of an overload, thus  facilitating the deployment of operators in a heterogeneous and/or resource-limited environment. 

\end{abstract}


\copyrightnotice

\section{Introduction} \label{sec:introduction}

The advent of event-driven applications and new stream processing technologies has facilitated
the creation and deployment of effective stream execution environments that gather and process large 
streams of data in distributed systems. 
Complex event processing (CEP) systems are therefore widely used in applications such as stock market surveillance, traffic monitoring and online fraud detection in banking systems.

In general, CEP systems enable timely correlation of data within distributed environments by defining specific groups of events or \textbf{patterns} to be detected together in a specific time window. Their usage 
is particularly relevant in situations in which multiple sources generate events which must be analyzed in real-time and in the 
same context, where the order of the detection is important. By distributing event detection across different computing units in a distributed environment, it is possible to parallelize the detection of 
specific parts of a complex event pattern effectively across multiple devices.

The challenge of deploying pattern detecting processes or operators to  specific nodes in a distributed environment arises when the resources
available to the individual computing units differ too greatly or are too restricted. The workload distribution
in a CEP system is generally uneven, operators effectively act as event filters, selectively processing events that match defined patterns, so that the total amount of events in circulation decreases further away from event sources. Additionally, 
the production rates of events might also fluctuate, causing a busy operator to spontaneously become idle or 
vice versa. This might lead to suboptimal utilization of the resources in the distributed environment, 
potentially causing operators with constrained processing capacity to overload, leading to queuing of events and slowing down pattern detection.

In many modern CEP systems the solution for the fluctuation of resource usage is addressed by the dynamic 
allocation of processing units in accordance with each operator's necessities~\cite{mayer_predictable_2015, lohrmann_elastic_2015, roger2019combining, russo_reinforcement_2019}. Once an operator becomes overloaded, 
its state is replicated in a newly allocated computing unit and the workload is then shared between them. 
Analogously, the number of operator replicas can be reduced if the processing rates of fewer operators suffice. 
This approach, while effective, cannot be fully utilized in a heterogeneous or restricted distributed system, as it 
is not guaranteed that the available nodes can meet the processing requirements of an overloaded 
operator and/or are sufficiently easily reachable, so that splitting the operator's work does not introduce new
latency problems.

Alternatively, there is also the possibility of regulating the input rates of individual operators by means of 
dropping events, a technique called load shedding. Several works introduce
methods to execute load shedding locally in an operator, by determining the match probability of individual 
events~\cite{slo_pspice_2019,slo2019espice, SloBR22,slo2023gspice, zhao_load_2020}. 
These works focus on single operator systems and establish a foundation for load regulation of individual operators 
by excluding events that are less likely to match patterns. 
While effective in the local operator scale, increasing the output of a single operator, they might not necessarily lead to a higher total system output
and can be complemented to consider the global relevance of events in a distributed application when determining their overall importance. 

Though there are many stream processing solutions which can be used to flexibly create and manage operator 
processes on independent hosts, there is a lack of open-source, easily deployable frameworks which offer sufficient load regulation control for both local and global processing scopes 
in a multi-operator CEP system.  As a result, our goal is to fill this gap by 
introducing an extensible CEP framework, Svayam, for traditional stream processing systems focused on 
fine-grained load control of individual operators in a distributed heterogeneous environment.

Svayam defines a set of basic components for the core operator processing logic and additionally introduces monitoring tools for continuous surveillance of an operator's stream characteristics. It enables the system to actively control which patterns are being detected by which operator and monitor the processing efficiency of those patterns in real time during execution. The main objective of Svayam is to provide a self-adaptive mechanism in case of operator overload, where the operator receives more events than it can process within the application's acceptable latency bounds. By employing load shedding, it determines the events that the operator processes while dropping remaining events from the input stream entirely, thus reducing the total average processing time of the operator significantly. Events are not dropped randomly, their importance to each event pattern of an operator is first evaluated and then measured with the necessity of shedding. Svayam components aim therefore at enforcing processing and efficiency requirements
for the Service-Level Objectives \cite{CasamayorPujolDMMD23} of operators in the edge 
continuum, guaranteeing system-wide operability even in resource-restricted 
environments.

More specifically, our contributions in this paper include a distributed extensible multi-operator CEP framework that detects overload and adapts itself by shedding load under the three following modes:
\begin{enumerate}
    \item Local mode: calculates the direct relevance of an event to the operator processing it, thus maximizing the local operator output.
    \item Global mode: leverages the global importance of an event to the total output of the system, considering all other complex events that could derive from it.
    \item Hybrid mode: combines both approaches, calculating the local relevance of the event and enriching it with its global relevance.
\end{enumerate}
Additionally, we have released the source code of Svayam as an open source project\footnote{https://github.com/O-Link06/Svayam}.


\section{Background} \label{sec:System Model}
\subsection{Events}
In a framework focused on data processing, the most fundamental abstraction is the one used for the 
representation of data itself. In the case of Svayam, this abstraction takes the form of an object 
class called Event. It represents the measurement of information at a source and serves as a 
primitive or complex event to be processed. 
Formally, we define an event $e_i$ as a tuple: $e_i = (T, ts, id)$, where:
\textbf{(i)} type ($T\in \mathbb T)$: the event type serves to differentiate various events, e.g., in a stock market application, it could represent a stock symbol, while in a transportation application, it could represent a specific bus ID,
\textbf{(ii)} timestamp (ts): represents the time of creation of an event in the system, 
\textbf{(iii)} id: a uniquely identifying number assigned to each event by the system.

\subsection{Stream Execution Environment}
In order to submit a processing job to a CEP system, we must first formalize which event 
types are to be detected together, the maximum allowed time difference between those events and 
which kinds of complex events are created once a match has been found. 
This information is represented in a structured request called the application query and it is used to 
create the operators ($\omega\in\Omega$), which together 
try to find groups of primitive events matching the query's specification. 

A single operator alone can correlate the information of multiple sources, but 
the real strength of CEP really becomes apparent when multiple operators work together 
to detect complex event patterns. Interconnected operators form an operator 
graph, linked to both the event generation applications as well as to the data consuming sinks responsible for ingesting the output of the CEP system. 

Together these components build the basic data transformation pipeline necessary for 
converting raw primitive events into semantically enriched complex events. Although 
their creation and execution are independent, they are all bound together by a common  
connecting layer, the stream execution environment. Created by a stream 
processing system, it provides and manages the basic infrastructure for the exchange of 
events between processes in the operator graph in continuous data streams. In this context, the system's job is to determine the number of operators necessary 
to implement a processing query's detection task, to create those operators and then 
use the overlaying stream execution environment to deploy them.

\subsection{Event Patterns}
Formally, an event pattern ($\gamma\in\Gamma$) represents a combination of event types, possibly ordered or unordered. The application query defines and associates each $\gamma$ with a corresponding complex event type representing its detection within a specific time window (difference of time between first and last events). 
An operator might be assigned one or multiple patterns and is then responsible for keeping track of the events it consumes, trying to find event types that together correspond to its assigned patterns and creating complex events once matches have been detected. We use $\Gamma_\omega$ to refer to the set of all patterns processed by an operator.
Event patterns can be of various categories, the most representative ones are as follows:
    (i) AND: unordered; all event types of the combination must be detected at least once for a match.
    (ii) OR: unordered; a match is produced when at least one event type has been detected, regardless of order.
    (iii) SEQ: strictly ordered; defines a sequence of event types that must be detected for a match. An event type $T$ might appear any number of times in the sequence.

    
\section{The Svayam Framework} \label{sec:framework}

\subsection{System Model}
\subsubsection{Pattern Detection with Finite State Machines}
Once a pattern $\gamma$ has been defined and deployed to an operator, it is then the framework's task to attempt to find 
events that follow the exact typing and timing constraints 
specified by it. In order to achieve this, an operator uses Finite State Machines or FSMs to model the 
different detection steps necessary for a successful match. By this method, each possible partial 
match leading to the full detection of $\gamma$ is modeled as a state in the 
Finite State Machine and each incoming event advances this state further, bringing it closer to 
the end state which represents pattern detection and leads to the creation of a new complex event once it has been reached. 

Maintaining a single Finite State Machine to process all events in the input stream would be 
counter-productive and does not fully capture the variety of different event combinations possible in a real-world application. Therefore, operators use a list of FSMs per pattern instead and thus evaluate an 
incoming event in multiple possible matching contexts. 
The FSM-lists of all operator-patterns are gathered and managed together by an FSM-Engine. The engine is responsible for iterating through the lists of all patterns sequentially for all incoming events, checking for matches and producing complex events or advanced FSMs accordingly.

Our default engine follows an exactly-once consumption policy. This means that, even though an event is used to advance the state of multiple FSMs, it can only produce at 
most one match. If an event can match multiple FSMs in the list, then it will select the one with the oldest timestamp within $\gamma$'s time window, this is referred to as a single, oldest-first selection-policy. Please note that the engine can be easily modified to support other selection and consumption policies.

\subsubsection{Quality and Latency Bound}

The quality of a CEP system is defined by the amount of complex events correctly 
identified for a given stream of primitive events. In an ideal scenario, all possible 
patterns are detected as soon as possible, ensuring that complex events are produced in a timely manner and
in a correct sequential order.

In a real scenario this is not always guaranteed, as it requires that each operator 
consistently works efficiently and that the communication between the components of the 
system is always timely. Thus we can only approximate a real system's quality value by 
determining how much the ideal ($\mathbb O_{out}^*$) and real ($\mathbb O_{out}$) outputs differ from each other. This 
difference, also called recall, is measured as the ratio of events produced in the real 
scenario compared to the ideal amount of events $r = \mathbb O_{out}/\mathbb O_{out}^*$.

CEP applications typically impose a latency bound ($L_{max}$) on an operator's processing time ($ptime_\omega$) to ensure real-time execution. However, fluctuating event production rates make maintaining this constraint challenging, especially as the event input rates ($\lambda_{in,\omega}$) of an operator exceed its processing rates ($\mu_\omega$). Consequently, load 
shedding becomes necessary, preemptively reducing $ptime_\omega$ and ensuring that it remains within $L_{max}$.
This inherently reduces system quality, as potentially matching events may be 
dropped, thus creating the need for a framework that maximizes the system's recall when 
determining an operator's shedding configuration.

\begin{figure*}[t!]
\centerline{\includegraphics[width=\textwidth]{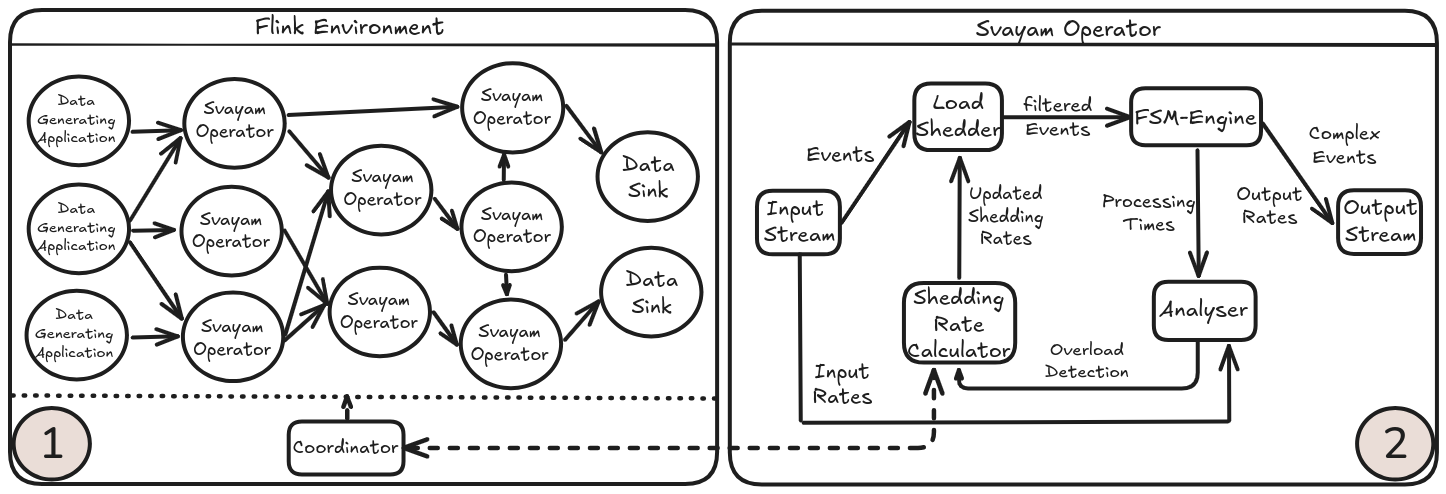}}
\caption{ The Svayam Framework - \textbf{A:} Example of an operator graph as seen in a distributed environment. 
\textbf{B:} Internal components of the 
{Svayam}
operator.
}
\label{fig:DiEnv}
\vspace{-8pt}
\end{figure*}

\subsection{Framework Architecture} \label{sec:System Architecture}

Svayam's CEP environment includes four key elements: \textbf{(i)} data generating applications, \textbf{(ii)} 
Svayam
operator,
\textbf{(iii)} data sinks, and \textbf{(iv)} a coordinator for global operator graph management (Figure \ref{fig:DiEnv} - A). Data generating applications and data sinks are external, while operator 
and the coordinator are core components.

Svayam Operators
expand the functionality of simple event processing operators $\omega$ to 
include monitoring of stream characteristics and load regulatory capabilities, so that they can dynamically adapt to changes of their input rates during execution. 
The operators contain: 
an analyzer, 
an FSM-Engine,
a shedding rate calculator and
a load shedder (Figure \ref{fig:DiEnv} - B).
The analyzer monitors $\lambda_{in,\omega}$, $ptime_\omega$ and $\mu_\omega$, detecting overloads. Upon overload, the shedding rate calculator computes
a new shedding rate configuration ($c_\omega$),  assigning shedding shares ($x_{\gamma,T}\in[0,1]$) to each event type $T$ in each pattern $\gamma\in\Gamma_\omega$. The load shedder receives $c_\omega$ and drops events from the input stream with probability $x_{\gamma,T}$ before they are processed by the different $\gamma$-FSM-engines.

Shedding rates can be calculated locally, prioritizing frequent event types in slow patterns, or globally by the coordinator. The coordinator is a centralized component that gathers operator information to assess impact on system output $\mathbb O_{out}$, allowing operators to shed events with low global contribution. The \textbf{Shedding Rate Calculator} component is abstract and represents both local and global shedding calculation options.

\subsection{Overload Detection} \label{sec:Overload Detection}
Correctly determining an operator overload and calculating an efficient load shedding 
configuration are essential for Svayam's load regulatory mechanisms and are 
intrinsic to the operator's
operations. Therefore, we discuss these in detail later in this section, specifying the shedding approaches in the three different modes.
The detection of an operator overload  is the same across all three different approaches. 
To have sufficiently fine-grained information, each operator continuously measures $\mu_\omega$, $\lambda_{in,\omega}$ and its output rates ($\lambda_{out,\omega}$) with an event-type precision. It also additionally measures the processing time of events in each pattern $ptime_{\gamma,\omega}$. These measurements are aggregated using running averages of adjustable size and 
are made available to the Analyzers each time they are updated.

An Analyzer requires both $\lambda_{in,\omega}$ of each input event type as well as $ptime_{\gamma,\omega}$ to determine an operator overload. 
To that end, it approximates 
the operator's total average processing time by correlating the proportion of events of 
each input type to the processing time of each pattern. This allows it to estimate the time required to completely process a single event. Once this measurement reaches or 
surpasses the latency bound, an overload is detected and a warning signal is 
produced.

Analyzers use the upstream operators' output rates as input rates, since locally measured input rates would not accurately represent an overloaded operator's real workload.

\subsection{Shedding Modes} \label{sec:Shedding Modes}
\textbf{Local Mode:} 
This mode focuses on determining a new load shedding configuration $c_\omega$ as quickly as 
possible, using only the local context of the operator, that is, the shedding rate 
calculator utilizes only the information it can locally collect in the {operator,}
thus reducing the necessary communication between operators.
Svayam provides a default implementation, in which the operators directly 
calculate $x_{\gamma,T}$ for all event types $T$ and all patterns $\gamma$ by measuring their 
proportional importance in $\lambda_{in,\omega}$ and $\lambda_{out,\omega}$ respectively and 
scaling this value with a correction factor defined as the ratio of the total measured processing time and the latency bound ($\frac{ptime_\omega}{L_{max}}$).
This approach can be changed or extended accordingly with other local algorithms focused on different stream characteristics or heuristics.

\textbf{Global Mode:}
Unlike the local mode, the global mode fully utilizes the enriched 
information available by using the Coordinator. Through its connection
with all operators in the operator graph, the Coordinator gathers 
the measured rate metrics $\lambda_{in,\omega}$, $\lambda_{out,\omega}$ and $\mu_{\omega}$ as well as $ptime_{\omega}$ from all operators $\omega$ in $\Omega$, allowing it to determine the frequency of event production and processing across the system. It can also assess how these events ultimately affect $\mathbb O_{out}$. 

Once an Analyzer detects an overload, it then requests the gathering of global 
information by the Coordinator, so that it can be used to determine the shedding 
configuration $c_\omega$ that maximizes $\mathbb O_{out}$ while reducing $ptime_\omega$.
In Svayam's default implementation the calculation of the new shedding rates is 
done directly at the Coordinator using a linear solver. By correlating variable 
shedding rates at the overloaded operator $\omega_o$ with the global information collected from the system, we can recursively determine which event types were more useful in that global state. Using that, it is possible to set constraints 
that limit the outputs of $\omega_o$ to match the real output of the 
system. That is, events of patterns that were unnecessarily produced and which ultimately 
contributed less to $\mathbb O_{out}$ are assigned higher shedding rates, considering 
also the specific processing time of each pattern 
$ptime_{\gamma,\omega_o}$.

Ultimately, to maximize $\mathbb O_{out}$ while minimizing $ptime_\omega$, the solver prioritizes event types that yield the highest number of relevant global matches with the least processing time.

\textbf{Hybrid Mode:}
Although the global mode directly benefits from the enriched information of the 
Coordinator, it lacks the speed of the local mode, as operators
must wait for the global information to be gathered and processed before  
being able to calculate their shedding rates.
In order to bridge the gaps of both methods, Svayam provides a default hybrid 
approach uniting their strengths.

If an operator $\omega$ overloads, its Analyzer will trigger both the calculation of a local shedding configuration $c_l$, prioritizing its direct outputs, as well as signal the Coordinator and request global information as in the Global Mode. Once the global
shedding configuration $c_g$ arrives, it substitutes $c_l$ and the operator starts 
prioritizing events $e$ with global relevance.

This approach is particularly useful when the input rates change rapidly and the 
communication with the Coordinator is slow. This prevents 
the operator from further overloading with increasing inputs while waiting for 
$c_g$.

\section{Evaluations} \label{sec:Evaluations}
\subsection{Experimental Setup} \label{sec:Experimental Setup}
In order to demonstrate the efficiency and latency properties of each mode we 
provide an example deployment of Svayam's framework in an Apache Flink Stream 
Execution Environment. We deploy the 
{operators}
and the coordinator using the 
ProcessFunction interfaces of Flink and connect them using 
DataStream objects and Apache Kafka Connectors for message exchange between 
components.
To facilitate this process we automate creation of
the operator graph and standardize the utilization and parsing of a query. 
Though our example focuses on Apache Flink, Svayam's components can be used 
independent of the stream execution environment.

We used artificial data based on the historical stock dataset of the S\&P 500 index in 
the years 2013 until 2018 for our analysis, available at \cite{dataset}. 
It includes records of the format: 
$stock = <date,open,high,low,close,volume,symbol>$
from which we extract events with the format:
$e=<symbol[0].up,timestamp,id>$
when the open is greater than the close 
value or using $symbol[0].down$ otherwise, whereby symbol[0] 
represents the first letter of the stock symbol. 

This way we group stocks with the 
same starting letter together and can track whether their prices more frequently 
increased or decreased through the measurements or how they can possibly 
correlate. Specifically, we aim to detect price oscillation waves (a stock transitions between up and down) and clashes (stocks 
of two different groups go up and one of them goes down in any order). 

The synthetic dataset is composed of 340,000 records divided in two files with 170,000 records each. We use two source components to read 700 records per second, simulating a constant stock price surveillance. We set a latency bound of $6\times 10^{-4}s$, slightly lower than the average record production time of $1.43\times 10^{-3}s \simeq 1/(700 eps)$.
Two operators $\omega_1$ and $\omega_2$ are connected to the sources and analyze clashes and waves for the stock groups A/G and C/L respectively. 
Next, operators $\omega_3$ and $\omega_4$ attempt to match clash and wave events coming from both upstream operators. In this example the combined outputs of $\omega_3$ and $\omega_4$ form $\mathbb O_{out}$. Below we show a formal representation of the query used, all patterns have a time window of $5s$: 

\noindent\textbf{Operator $\omega_1:$}

\noindent\textbf{Select:} $G.up, G.down, A.up, A.down$ from Source1

\noindent\textbf{Where:}
\textit{P1:}$AND_{5s}(G.up,A.up,G.down)$ as $clash.GA$

\noindent\textit{P2:} $SEQ_{5s}(G.down,G.up,G.down)$ as $wave.G$

\noindent\textit{P3:} $SEQ_{5s}(A.up,A.down,A.up)$ as $wave.A$

\noindent\textbf{Operator $\omega_2:$ }

\noindent\textbf{Select:} $L.up, L.down, C.up, C.down$ from Source2

\noindent\textbf{Where:}
\textit{P1:} $AND_{5s}(C.up,C.up,L.down)$ as $clash.CL$

\noindent\textit{P2:} $SEQ_{5s}(L.down,L.up,L.down)$ as $wave.L$

\noindent\textit{P3:} $SEQ_{5s}(C.up,C.down,C.up)$ as $wave.C$

\noindent\textbf{Operator $\omega_3:$}

\noindent\textbf{Select:} $clash.CL, clash.GA$

\noindent\textbf{Where:}
\textit{P1:} $AND_{5s}(clash.CL, clash.GA)$ as $clash.cx$

\noindent\textbf{Operator $\omega_4:$ }

\noindent\textbf{Select:} $wave.G, wave.A, wave.L, wave.C$

\noindent\textbf{Where: }
\textit{P1:} $AND_{5s}(wave.A,wave.C)$ as $wave.cx1$

\noindent\textit{P2:} $AND_{5s}(wave.L,wave.G)$ as $wave.cx2$


\subsection{Results} \label{sec:Result}
\begin{figure}[htbp] \centerline{\includegraphics[width=\linewidth]{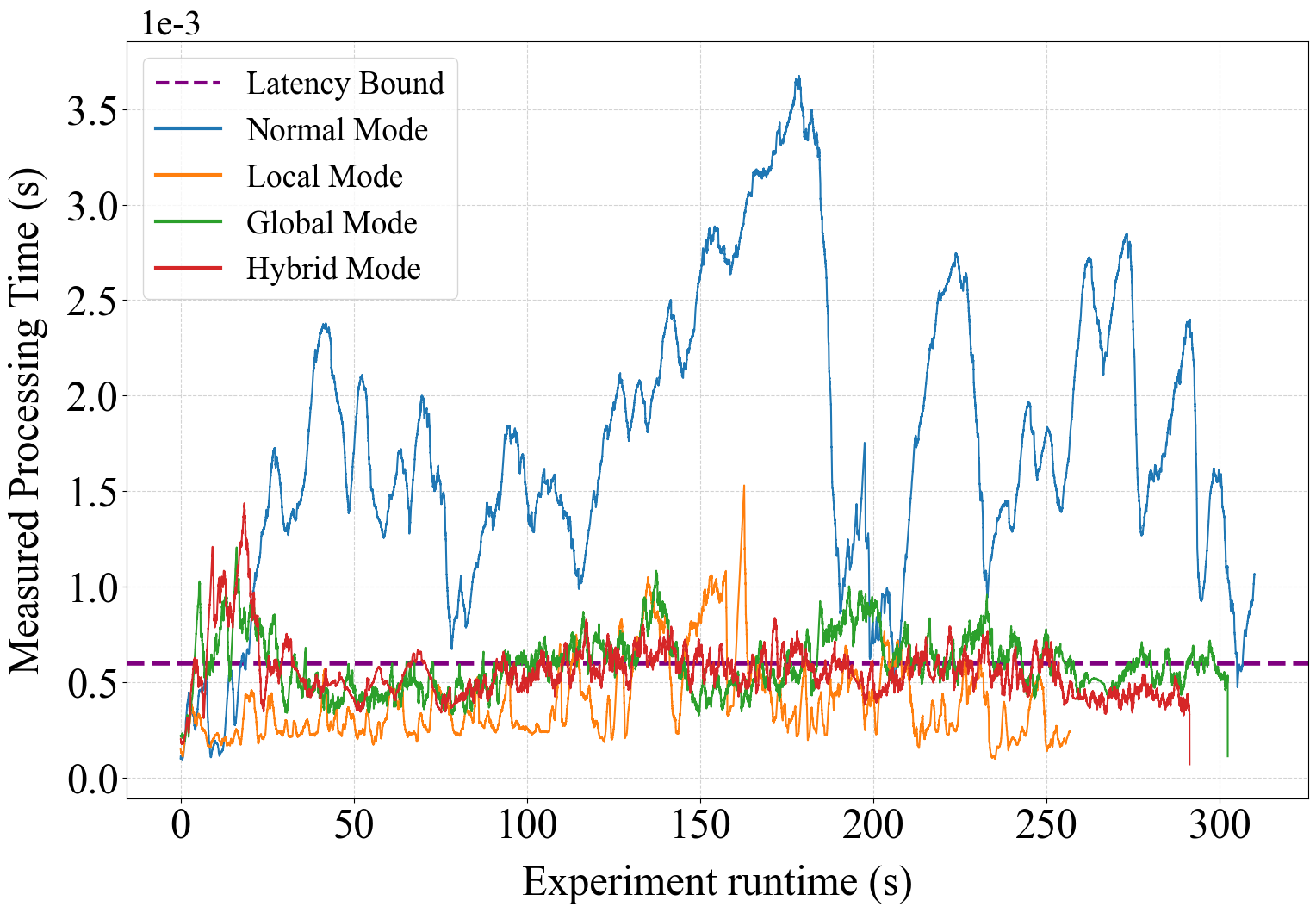}}
\vspace{-8pt}
\caption{Latency compliance of the different modes.}
\vspace{-8pt}
\label{fig:graph}
\end{figure}

\begin{figure}[hbtp]
\centerline{\includegraphics[width=160pt]{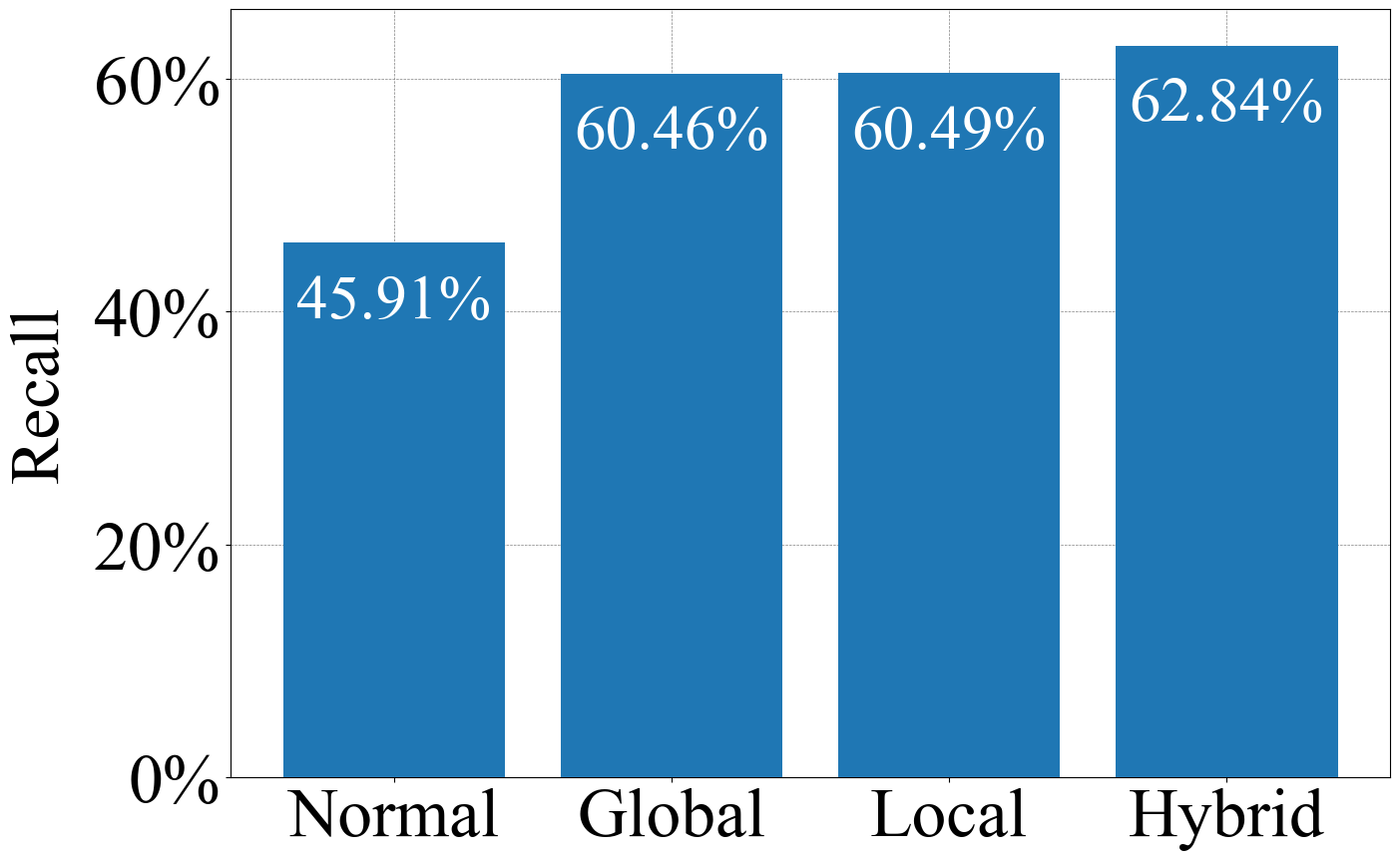}}
\vspace{-8pt}
\caption{Comparison of the recall in the different modes.}
\label{fig:recall}
\vspace{-8pt}
\end{figure}
Tests were conducted on a standalone, containerized Flink cluster using an Intel Core i7-2670QM Processor with 16 GB DDR3 RAM. 
For comparative analysis, we implemented an additional \textbf{Normal Mode} without 
load shedding and compared it with our three approaches (Local, Global and 
Hybrid). 
In our tests we analyzed both the latency compliance of the different modes with 
respect to the given latency bound (Figure \ref{fig:graph}) as well as their recall (Figure \ref{fig:recall}). To estimate the maximum achievable system output ($\mathbb O_{out}^*$) under ideal conditions, we processed each operator individually without load shedding and with synchronized input timestamps and ideal processing/exchange times to determine their optimal outputs and detect all possible matches within their time windows.

As expected, the normal mode without load shedding exhibited the highest processing time. Operators which could not process their records in time blocked their communication channel, stalling the production of new complex events. Consequently the total experiment runtime (557 s) was considerably higher than the combined average of the 3 other modes (287 s) and was shortened for visibility in Figure \ref{fig:graph}. Local, global and hybrid mode all achieved a substantial reduction in processing time compared to the normal mode, with most of their data points consistently falling below the latency bound and quickly adapting to increasing processing times.

Another major impact of load shedding is visible in Figure \ref{fig:recall}. Despite processing all incoming events, the normal mode had the lowest recall value (45\%). This counterintuitive result is attributed to the difference in speed between operators. Without load shedding, faster operators can overwhelm slower ones, leading to complex events being generated out of order or with significant delays. Consequently, crucial event correlations are missed, resulting in a substantial drop in recall.
In contrast, all load shedding modes achieved significantly higher recalls (ranging from 60.46\% to 62.83\%) by strategically reducing the workload of their operators. Enforcing a common latency bound helped these modes maintain more consistent processing speeds across the operators, ensuring that matching events produced by different operators were more likely to be generated and detected together within the same time window. 

Although our three modes demonstrated similar recall results for this query in our test environment, they each introduce unique nuances to the shedding process, offering different trade-offs in terms of information availability, decision-making complexity and operational overhead. It is therefore paramount to carefully consider an appropriate shedding strategy based on factors such as deployment complexity, resource requirements and sensitivity to dynamic changes in the workload.

\section{Related Work}

While there are load shedding solutions available for single-node applications \cite{slo2019espice, slo_pspice_2019, katsipoulakis2018concept, conf/debs/SloBR20}, few are available for the introduced distributed setting. 
Load shedding in stream processing, as explored by Katsipoulakis et al.\cite{katsipoulakis2018concept}, adapts to dynamic input-stream distributions by frequently sampling from arriving windows while focusing on a single operator. Their work supports the notion that dynamic selectivities are common in stream processing and can be leveraged, particularly in join operators and aggregations.
Tatbul et al. \cite{tatbul_staying_2007} explore load shedding for single and multiple operators, leveraging selectivity functions for pre-processing. They formulate the problem as a linear optimization task and propose a solver-based centralized approach and a distributed approach that rely on precomputed load shedding plans.
However, these stream processing approaches do not account for the complexities of CEP operators, i.e., conditional operators in distributed CEP.

Slo et al. \cite{slo2019espice, slo_pspice_2019, SloBR22, slo2023gspice} focus on load shedding with a focus on pattern detection.  
They learn an event's utility based on its context, type, or the operator's internal state. 
However, in a multi-operator scenario, a shedder cannot tell at which position output events of its operators can end up for downstream operators.
For single and even complex operators, these are useful solutions. However, in a distributed setting, the used utility values might fall short as introduced before. 

 In approximate processing  only a share of the input stream is processed. 
While the perspective focuses on what to process, not on what to drop in approximate processing, the outcome is similar. However, approximate processing is primarily used for stream processing and not CEP\cite{FarhatBD20, GEDIK2016106}. 

{Additionally, Choochotkaew et al. propose an edge-adaptive CEP system \cite{8271954} that dynamically changes operator detection tasks based on resource availability and proximity to sources and sinks, effectively collecting edge data, but limiting pattern complexity by not considering multi-operator complex processing pipelines as seen in traditional operator graphs.}


\section{Challenges and Future Work}
Creating and deploying Svayam in a distributed system highlighted several aspects worth further inspection:

{
\textbf{1) Dynamically tuning hyperparameters:} 
In distributed systems, varying operator complexity (multiple patterns or high message congestion) complicates defining latency bounds and control batch sizes for stream measurement, which are crucial for timely overload response. Svayam currently uses fixed values for those control parameters, but future versions should dynamically adapt them to runtime input rate fluctuations.


}

{
\textbf{2) Implementing additional pattern categories, quantifiers and parameterization:}
Svayam currently supports basic AND, OR and SEQ operators with a limited number of event types per pattern. 
Future development should include the NOT operator and type quantifiers for potentially
infinite event sequences. Additionally, Svayam should also support event parametrization, 
increasing event pattern complexity.
}

{
\textbf{3) Optimization of data structures:} 
Despite careful consideration, Svayam's data structures are not fully optimized for inter-operator processing and (de-)serialization time; future work includes optimizing these structures and their serialization using Apache Avro for better system integration.
}

\textbf{4) Generalization of components:} Currently the abstractions used for creating different pattern categories and operator types are non-intuitive. The goal of the next optimizations should be organizing these abstractions with well known design patterns like factory or prototype.
\section{Conclusion}
In a heterogeneous CEP application distributed across the cloud edge continuum, individual operators can become bottlenecks, increasing end-to-end latency. When resources are limited, load shedding is an effective strategy to maintain timely processing of critical events while discarding less important ones. As a result, in this paper, we presented Svayam, a self-adaptive extensible CEP framework that  enables load shedding in geographically distributed CEP applications by introducing conditional selectivity functions for key building blocks of CEP queries. Svayam can be used in three different shedding modes, whereby it  detects overload over a multi-operator environment and successfully adapts itself to meet an application's  latency requirements. 

\bibliographystyle{plain}
\bibliography{references}

@misc{dataset,
  author = {{Cam Nugent}},
  title = {Historical stock data for all current S\&P 500 companies},
  howpublished = {\href{https://www.kaggle.com/datasets/camnugent/sandp500}},
  note = {[Accessed Online: 2025-03-01]} 
}

@INPROCEEDINGS{8271954,
  author={Choochotkaew, Sunyanan and Yamaguchi, Hirozumi and Higashino, Teruo and Shibuya, Megumi and Hasegawa, Teruyuki},
  booktitle={Proc. of Int. Conf. on Distributed Computing in Sensor Systems}, 
  title={EdgeCEP: Fully-Distributed Complex Event Processing on IoT Edges}, 
  year={2017},
  pages={121-129}
  }

@article{GEDIK2016106,
title = {Pipelined fission for stream programs with dynamic selectivity and partitioned state},
journal = {Journal of Parallel and Distributed Computing},
volume = {96},
pages = {106-120},
year = {2016},
issn = {0743-7315},
doi = {https://doi.org/10.1016/j.jpdc.2016.05.003},
url = {https://www.sciencedirect.com/science/article/pii/S0743731516300338},
author = {B. Gedik and H.G. Özsema and Ö. Öztürk}
}

@inproceedings{roger2019combining,
  title={Combining it all: Cost minimal and low-latency stream processing across distributed heterogeneous infrastructures},
  author={R{\"o}ger, Henriette and Bhowmik, Sukanya and Rothermel, Kurt},
  booktitle={Proc. of the 20th International Middleware Conference},
  pages={255--267},
  year={2019}
}

@inproceedings{russo_reinforcement_2019,
	title = {Reinforcement Learning Based Policies for Elastic Stream Processing on Heterogeneous Resources},
	isbn = {978-1-4503-6794-3},
	url = {http://doi.acm.org/10.1145/3328905.3329506},
	doi = {10.1145/3328905.3329506},
	booktitle = {Proc. of the 13th {ACM} Int. Conf. on Distributed and Event-based Systems},
	author = {Russo, Gabriele Russo and Cardellini, Valeria and Presti, Francesco Lo},
	urldate = {2019-08-28},
	year = {2019}	
}

@inproceedings{lohrmann_elastic_2015,
	title = {Elastic Stream Processing with Latency Guarantees},
	doi = {10.1109/ICDCS.2015.48},
	eventtitle = {2015 {IEEE} 35th International Conference on Distributed Computing Systems},
	pages = {399--410},
	booktitle = { {IEEE} 35th Int. Conf. on Distributed Computing Systems, {ICDCS}},
	author = {Lohrmann, B. and Janacik, P. and Kao, O.},
	year = {2015}
}

@article{mayer_predictable_2015,
  author       = {Ruben Mayer and
                  Boris Koldehofe and
                  Kurt Rothermel},
  title        = {Predictable Low-Latency Event Detection With Parallel Complex Event
                  Processing},
  journal      = {{IEEE} Internet Things Journal},
  volume       = {2},
  number       = {4},
  pages        = {274--286},
  year         = {2015},
  url          = {https://doi.org/10.1109/JIOT.2015.2397316},
  doi          = {10.1109/JIOT.2015.2397316},
  bibsource    = {dblp computer science bibliography, https://dblp.org}
}

@inproceedings{slo_pspice_2019,
  author       = {Ahmad Slo and
                  Sukanya Bhowmik and
                  Albert Flaig and
                  Kurt Rothermel},
 
  title        = {pSPICE: Partial Match Shedding for Complex Event Processing},
  booktitle    = { {IEEE} Int. Conf. on Big Data (BigData)},
  year         = {2019}
}

@inproceedings{slo2019espice,
  title={{espice}: Probabilistic load shedding from input event streams in complex event processing},
  author={Slo, Ahmad and Bhowmik, Sukanya and Rothermel, Kurt},
  booktitle={Proc. of the 20th Int. Middleware Conference},
  pages={215--227},
  year={2019}
}

@inproceedings{zhao_load_2020,
	location = {Dallas, {TX}, {USA}},
	title = {Load Shedding for Complex Event Processing: Input-based and State-based Techniques},
	isbn = {978-1-72812-903-7},
	url = {https://ieeexplore.ieee.org/document/9101376/},
	doi = {10.1109/ICDE48307.2020.00099},
	shorttitle = {Load Shedding for Complex Event Processing},
	eventtitle = {{IEEE} 36th Int. Conf. on Data Engineering ({ICDE})},
	pages = {1093--1104},
	booktitle = { Proc. of the {IEEE} 36th Int. Conf. on Data Engineering ({ICDE})},
	publisher = {{IEEE}},
	author = {Zhao, Bo and Viet Hung, Nguyen Quoc and Weidlich, Matthias},
	urldate = {2020-12-08},
	year = {2020},
	langid = {english},
}

@inproceedings{tatbul_staying_2007,
	location = {Vienna, Austria},
	title = {Staying {FIT}: Efficient Load Shedding Techniques for Distributed Stream Processing},
	isbn = {978-1-59593-649-3},
	url = {http://dl.acm.org/citation.cfm?id=1325851.1325873},
	series = {{VLDB} '07},
	shorttitle = {Staying {FIT}},
	pages = {159--170},
	booktitle = {Proc. of the 33rd Int. Conf. on Very Large Data Bases},
	publisher = {{VLDB} Endowment},
	author = {Tatbul, Nesime and Çetintemel, Uǧur and Zdonik, Stan},
	year = {2007},
}

@inproceedings{katsipoulakis2018concept,
  title={Concept-driven load shedding: Reducing size and error of voluminous and variable data streams},
  author={Katsipoulakis, Nikos R and Labrinidis, Alexandros and Chrysanthis, Panos K},
  booktitle={ IEEE Int. Conf. on Big Data (Big Data)},
  pages={418--427},
  year={2018},
  organization={IEEE}
}

@article{SloBR22,
  author       = {Ahmad Slo and
                  Sukanya Bhowmik and
                  Kurt Rothermel},
  title        = {State-Aware Load Shedding From Input Event Streams in Complex Event
                  Processing},
  journal      = {{IEEE} Trans. Big Data},
  volume       = {8},
  number       = {5},
  pages        = {1340--1357},
  year         = {2022},
  url          = {https://doi.org/10.1109/TBDATA.2020.3047438},
  doi          = {10.1109/TBDATA.2020.3047438},
  bibsource    = {dblp computer science bibliography, https://dblp.org}
}

@inproceedings{conf/debs/SloBR20,
  author       = {Ahmad Slo and
                  Sukanya Bhowmik and
                  Kurt Rothermel},
  title        = {{hSPICE}: state-aware event shedding in complex event processing},
  booktitle    = {14th {ACM} Int. Conf. on Distributed and Event-based
                  Systems, {DEBS}},
  pages        = {109--120},
  publisher    = {{ACM}},
  year         = {2020},
  url          = {https://doi.org/10.1145/3401025.3401742},
  doi          = {10.1145/3401025.3401742},
  bibsource    = {dblp computer science bibliography, https://dblp.org}
}

@inproceedings{FarhatBD20,
  author       = {Omar Farhat and
                  Harsh Bindra and
                  Khuzaima Daudjee},
  title        = {Leaving stragglers at the window: low-latency stream sampling with
                  accuracy guarantees},
  booktitle    = {14th {ACM} Int. Conf. on Distributed and Event-based
                  Systems, {DEBS}},
  pages        = {15--26},
  publisher    = {{ACM}},
  year         = {2020},
}

@inproceedings{slo2023gspice,
  author       = {Ahmad Slo and
                  Sukanya Bhowmik and
                  Kurt Rothermel},
  title        = {gSPICE: Model-Based Event Shedding in Complex Event Processing},
  booktitle    = {{IEEE} Int. Conf. on Big Data (BigData)},
  pages        = {263--270},
  publisher    = {{IEEE}},
  year         = {2023},
  url          = {https://doi.org/10.1109/BigData59044.2023.10386775},
  doi          = {10.1109/BIGDATA59044.2023.10386775},
  bibsource    = {dblp computer science bibliography, https://dblp.org}
}

@article{CasamayorPujolDMMD23,
  author       = {Victor Casamayor{-}Pujol and
                  Praveen Kumar Donta and
                  Andrea Morichetta and
                  Ilir Murturi and
                  Schahram Dustdar},
  title        = {Edge Intelligence - Research Opportunities for Distributed Computing
                  Continuum Systems},
  journal      = {{IEEE} Internet Comput.},
  volume       = {27},
  number       = {4},
  pages        = {53--74},
  year         = {2023},
  url          = {https://doi.org/10.1109/MIC.2023.3284693},
  doi          = {10.1109/MIC.2023.3284693},
  bibsource    = {dblp computer science bibliography, https://dblp.org}
}

\end{document}